# Direct Data Domain STAP using Sparse Representation of Clutter Spectrum

Ke Sun[1], Huadong Meng[1], Yongliang Wang[2], Xiqin Wang[1]

([1]Department of Electronic Engineering, Tsinghua University, Beijing 100084, China

[2] Science and Research Department, Wuhan Radar Academy, Wuhan 430019, China)

**Abstract**: Space-time adaptive processing (STAP) is an effective tool for detecting a moving target in the airborne radar system. Due to the fast-changing clutter scenario and/or non side-looking configuration, the stationarity of the training data is destroyed such that the statistical-based methods suffer performance degradation. Direct data domain (D3) methods avoid non-stationary training data and can effectively suppress the clutter within the test cell. However, this benefit comes at the cost of a reduced system degree of freedom (DOF), which results in performance loss. In this paper, by exploiting the intrinsic sparsity of the spectral distribution, a new direct data domain approach using sparse representation (D3SR) is proposed, which seeks to estimate the high-resolution space-time spectrum with only the test cell. The simulation of both side-looking and non side-looking cases has illustrated the effectiveness of the D3SR spectrum estimation using focal underdetermined system solution (FOCUSS) and $L_1$ norm minimization. Then the clutter covariance matrix (CCM) and the corresponding adaptive filter can be effectively obtained. Since D3SR maintains the full system DOF, it can achieve better performance of output signal-clutter-ratio (SCR) and minimum detectable velocity (MDV) than current D3 methods, e.g., direct data domain least squares (D3LS). Thus D3SR is more effective against the range-dependent clutter and interference in the non-stationary clutter scenario.

## 1. Introduction

An airborne/spaceborne (A/S) space-time adaptive processor (STAP) attempts to detect a moving target in the presence of Doppler/angle spread clutter environment [1-5]. Due to the movement of the radar platform, neither one dimensional processing in the angle nor in the Doppler domain can effectively distinguish the moving target from the surrounding clutter environment. Therefore it is necessary to carry out the space-time joint processing. In the adaptive radar processing, system degree of freedom (DOF), (i.e., the number of independent space-time samples), indicates the adaptive filter's influence on the system performance. In a sense, it reflects the dimension of the total space provided by the STAP system. Pioneer works [1-2] have illustrated that the signal-clutter-ratio (SCR) is improved mostly in the subspace that is orthogonal to the clutter. When the dimension of the clutter subspace（i.e., the clutter rank）is fixed, a higher system DOF means a better SCR improvement by the STAP processor. Traditionally, statistical-based methods such as loaded sample matrix inversion (LSMI) and principle component (PC) [6-7] need statistically independent and identically distributed (IID) training data to obtain an effective estimation of the clutter covariance matrix (CCM), which can also be viewed as a representation of the clutter subspace. When they have sufficient IID training data, the adaptive filter can be constructed in such a way that the output SCR is greatly improved. However, if the clutter scenario is not homogeneous or the airborne system is deployed with a non side-looking and/or a conformal array radar, the range stationarity is destroyed, which results in an inaccurate estimation of the CCM and an improper nulling of the clutter [8-9].

A series of methods have been proposed to focus on the nonstationarity of the training data. Methods such as angle-Doppler compensation [10-11] and adaptive angle-Doppler compensation [12] accomplish the peak response, but the sidelobe clutter suppression is limited. The registration-based approach implements both the mainlobe and sidelobe clutter compensation [13-14]. However, the performance depends on the approximation between the registered spectrum and the actual spectrum. If there is a significant difference between the two spectra, which is common in the short-range case, the clutter cannot be sufficiently suppressed.

The direct data domain (D3) approach is proposed from a different perspective, which does not need training data [15-17]. This approach assumes that the direction and Doppler frequency of the assumed signal of interest (SOI) is known in advance. The goal of the D3 method is to search for a moving target located at a certain look direction and Doppler frequency. As D3 only uses the test cell but not the training data, it can avoids the nonstationarity in the training data and effectively suppress both the clutter and interference. However, there is a tradeoff between the system DOF of the adaptive filter and the number of the subarrays. Thus the benefit of not requiring the training data comes at the cost of a reduced system DOF, which results in decreased performance.

The key requirement for any form of STAP approach is the accurate knowledge of the clutter ridge in the test cell [3-4]. Once the clutter spectrum is obtained with high resolution, the corresponding adaptive filter can be constructed to improve the output SCR. Due to the system limitation in the airborne radar, the space-time samples, (i.e., the number of array channels and pulses), are insufficient. The slow-moving target is covered by the sidelobe of the surrounding clutter and not visible without the adaptive processing [1-2]. In this paper, a new D3 method, direct data domain via sparse representation (D3SR), is proposed, which seeks to obtain accurate clutter spectrum with only the test cell. D3SR converts the spectral estimation into the solution of the underdetermined inverse problem with sparse constraint. In its basic form, the technique of sparse representation attempts to find the sparsest signal $\boldsymbol{\alpha}$ to satisfy $\mathbf{x} = \boldsymbol{\Psi\alpha}$, where $\boldsymbol{\Psi} \in C^{m \times n}$ is an overcomplete basis, (i.e., $m \leq n$). Normally, the equation is ill-posed and has many solutions. Additional constraint that $\boldsymbol{\alpha}$ should be sparse allows one to eliminate this ill-posedness [18-19]. A number of practical algorithms such as $L_1$ norm minimization [20-21], and focal underdetermined system solution (FOCUSS) [22-23] have been proposed to approximate this sparse solution. Sparse representation has been illustrated as an effective tool in the spectrum estimation. However, the application has mainly been focused in the field of source localization and neuromagnetic imaging [24-26], where the sparsity is obvious.

In this paper, we exploit the priori sparsity of the spectral distribution in the test cell and propose a new approach to obtain the high-resolution spectrum using sparse representation, (which is developed from our earlier work [27]). Then accurate clutter distribution is extracted from this spectral result using the knowledge of the assumed SOI. Based on this, the CCM and the corresponding adaptive filter can be effectively obtained. D3SR avoids the training data and provides the adaptive filter with a higher system DOF than the current direct data domain least square (D3LS) method. Thus D3SR can effectively solve the problem of nonstationarity in both the side-looking and non side-looking cases and provide better performance of output SCR and minimum detectable velocity (MDV). The following parts of this paper are organized as follows. Section 2 describes the basic model of the test cell in STAP. Section 3 proposes D3SR to estimate the high-resolution space-time spectrum using sparse representation in both side-looking and non side-looking cases. Then the CCM and the corresponding adaptive filter are obtained to effectively suppress the clutter and interference. Section 4 uses the simulated data to test the performance of the SCR improvement and MDV. Section 5 gives a conclusion on the proposed algorithm and points out the future work.

## 2. Signal Model

In an actual airborne STAP system, the received data behave non-stationary due to many practical factors. In the side-looking case, the nonstationarity is mainly caused by the discrete interference and fast variation of the clutter scenario [2-4]. Because the scenario of interest is unknown in advance, this kind of nonstationarity is not predictable. Another important kind of nonstationarity is caused by the airborne radar array configuration [10-12]. For example, non side-looking array makes the clutter ridge range-dependent even when the scenario is fully homogenous. This nonstationarity is predictable to some extent if the geometry information is known from other equipments such as global position system or inertial navigation system.

The general geometry of an airborne radar array system is shown in Fig.1 [10], where the flight direction is along the $x$ axis with a velocity $v$, the vertical height of the platform is $H$, the array line is parallel to the XOY plane. Point $P$ stands for one scatter in the scenario of interest. The symbols $\theta$ and $\varphi$ are the elevation and azimuth angles, respectively, and $\psi$ denotes the crab angle between the array line and the flight direction. For example, $\psi = 0°, 90°$ denote the side-looking and forward-looking array orientations, respectively. Therefore, the Doppler frequency due to a certain stationary scatter is given as

$$f_d = \frac{2v}{\lambda} \cos\varphi \cos\theta, \qquad (1)$$

where the moving scatter case can be calculated similarly by adding its own velocity. The elevation angle $\theta$ can be expressed by the platform height $H$ and slant range $R_s$ as

$$f_d = \frac{2v}{\lambda} \cos\varphi \sqrt{1 - \frac{H^2}{R_s^2}}. \qquad (2)$$

Define $\beta$ as the radar look direction relative to the array line

$$\cos\beta = \cos(\varphi - \psi)\cos\theta. \qquad (3)$$

Thus, the Doppler frequency depends on the look direction and the crab angle as follows

$$f_d = \frac{2v}{\lambda}\left(\cos\psi \cos\beta \pm \sqrt{\cos^2\psi \cos^2\beta - \cos^2\beta + \sin^2\psi \cos^2\theta}\right), \qquad (4)$$

where the sign before the square root denotes the clutter Doppler frequencies left and right of the array axis. In the side-looking case of $\psi = 0$, which is discussed in fundamental STAP works, the Doppler and angle dependence is simplified as

$$f_d = \frac{2v}{\lambda}\cos\beta. \tag{5}$$

Consequently, the Doppler frequency only depends on the look direction $\beta$, and not on the range. Furthermore, when the scenario is fully homogenous, the training data behaves stationary and can be utilized to estimate the CCM so that the effective STAP filter is available to suppress the clutter in the test cell.

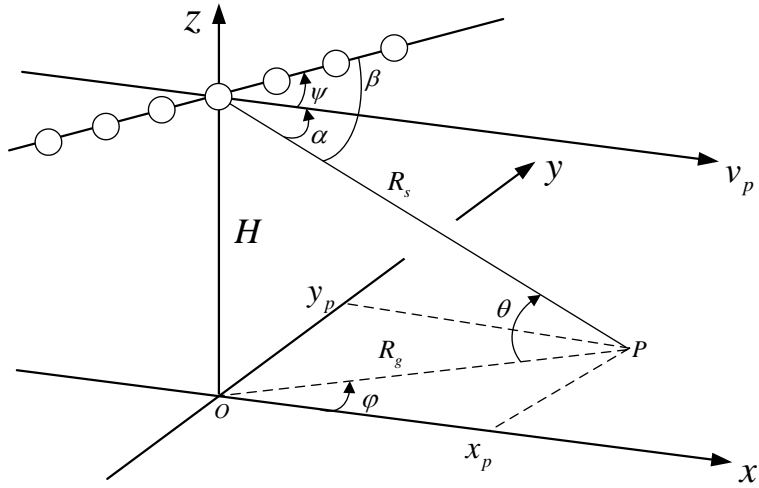

Fig.1 Geometry of a linear airborne array

As shown in Fig.2 (a) of the side-looking case, the clutter ridges of different range cells coincide with each other and behave a straight line in the angle-Doppler domain. Meanwhile, Fig.2 (b) shows that these trajectories change into a set of concentric ellipses instead of a straight line in the non side-looking case. Thus, the clutter ridge is range-dependent by the property of the airborne array configuration, and this nonstationarity of the training data becomes serious especially in the short-range scenario. The CCM estimation using the training data creates an improper filter and causes performance degradation. Next, we will briefly introduce the classic D3 model in both side-looking and non side-looking cases.

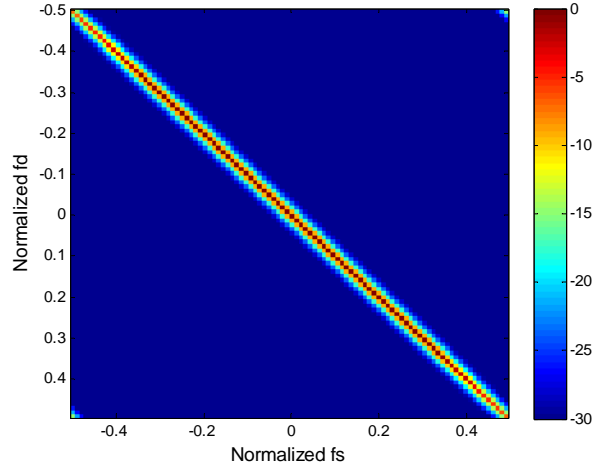

Fig.2 (a) Clutter ridges in side-looking case (dB)

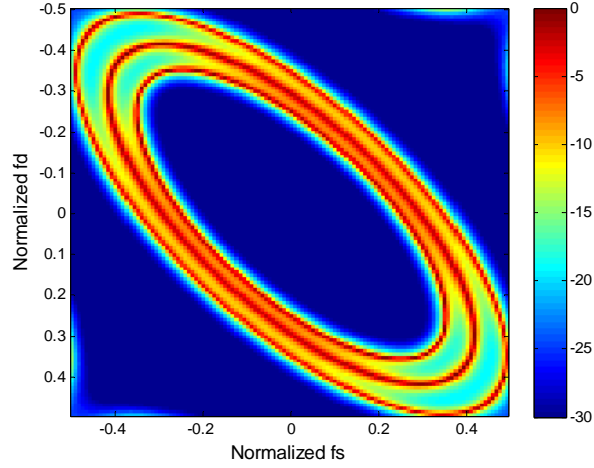

Fig.2 (b) Clutter ridges in non side-looking case (dB)

In the side-looking airborne radar, the classic non-stationary scenario is assumed to be composed with three components: the moving target, i.e., the signal of interest (SOI), the clutter return and discrete interferers [15-16]. The objective of STAP is to suppress both the clutter and interference while maintaining the system gain at SOI to improve the output SCR. Suppose that $N$ is the number of array channels and $M$ is the number of pulses in a coherent process interval (CPI). The space-time steering vector ($NM \times 1$) of the SOI component can be written as follows:

$$\mathbf{s} = \mathbf{a} \otimes \mathbf{b}, \qquad (6)$$

where $\otimes$ denotes the Kronecker product, $\mathbf{a}$ and $\mathbf{b}$ stand for the corresponding space and Doppler steering vectors, respectively as

$$\mathbf{a} = \left[1,\ \exp\left(j2\pi \frac{d}{\lambda}\sin\beta_s\right),\ldots,\exp\left(j2\pi(N-1)\frac{d}{\lambda}\sin\beta_s\right)\right]^T, \qquad (7)$$

and

$$\mathbf{b} = \left[1,\ \exp\left(j2\pi \frac{f_d}{\text{PRF}}\right),\ldots,\exp\left(j2\pi(M-1)\frac{f_d}{\text{PRF}}\right)\right]^T, \qquad (8)$$

where $d$ is the interelement spacing between the antenna elements, $\lambda$ is the wavelength of the radar, PRF is the radar pulse repetition frequency, $\beta_s$ and $f_d$ stand for the look direction and Doppler frequency of the SOI, respectively. Besides the SOI component, there is clutter, which is generated by the terrain scatterings and can be considered as a collection of independent scatters as

$$\mathbf{x}_c = \sum_{i=1}^{N_c} \gamma_i \cdot \mathbf{v}_i, \qquad (9)$$

where $N_c$ is the number of the statistical independent clutter scatters, $\gamma_i$ and $\mathbf{v}_i$ are the random complex amplitude and space-time steering vector of the $ith$ clutter scatter respectively. Due to the movement of the radar platform, the clutter has an angle-Doppler dependence in (5) [1-2]. The clutter is contained in both the training data and test cell, thus the statistical-based methods [6-7] can use the training data to obtain the clutter distribution and effectively suppress it in the test cell. In the actual scenario, there also exist some discrete interferers, which come from any stationary transmitter such as radio station [2-3]. The amplitude and Doppler frequency of the discrete interferers are often assumed to be stationary over the CPI. Consequently, the interference can be modeled in the same way as the clutter in (9) as follows:

$$\mathbf{x}_d = \sum_{l=1}^{N_d} \kappa_l \cdot \mathbf{d}_l, \qquad (10)$$

where $N_d$ is the number of discrete interferers, $\kappa_l$ and $\mathbf{d}_l$ are the random complex amplitude and space-time steering vector of the $lth$ discrete interferer, respectively. Unlike the clutter, the angle and Doppler frequency of the discrete interferers do not have a certain relationship as they do not appear along the clutter ridge. Using the above modeling, the $NM \times 1$ data of the test cell can be modeled as

$$\begin{aligned}\mathbf{x} &= \mathbf{x}_c + \mathbf{x}_d + \mathbf{x}_t + \mathbf{n} \\ &= \sum_{i=1}^{N_c} \gamma_i \cdot \mathbf{v}_i + \sum_{l=1}^{N_d} \kappa_l \cdot \mathbf{d}_l + \alpha \mathbf{s} + \mathbf{n},\end{aligned} \qquad (11)$$

where $\mathbf{x}_t$ is the component of the moving target and $\alpha$ is its complex amplitude. The additive component $\mathbf{n}$ is the thermal noise that is uncorrelated with the other components.

A series of D3 methods have been proposed that focus on the problem of nonstationarity [15-17]. The D3LS method assumes that the look direction and Doppler frequency of SOI is known in advance. If we define the subarray numbers along channel and pulse axes as $N_a, N_P$, which are determined by the system requirement, then the possible SOI can be cancelled from the test cell in the direct data domain to obtain a $(N_a N_P - 1) \times N_a N_P$ cancellation matrix $\mathbf{X}$, which does not contain any SOI component. The adaptive filter $\mathbf{w}$ that cancels both the clutter and interference while preserving the SOI can be found by solving the following equation:

$$\begin{bmatrix} \mathbf{s}_{sub} \\ \mathbf{X} \end{bmatrix} \mathbf{w} = \begin{bmatrix} C \\ 0 \\ \vdots \\ 0 \end{bmatrix}_{(N_a N_P \times 1)} \qquad (12)$$

where constant $C$ is the look-direction gain, and $N_a N_P \times 1$ vector $\mathbf{s}_{sub}$ is the subarray space-time steering vector of the SOI. The technique of least square is adopted here to solve equation (12) [15]. The D3LS method only uses the test cell and bypasses the problem of the required stationary training data. Thus it avoids the nonstationarity induced by the training data and can effectively suppress both the clutter and discrete interferers in the test cell. However this benefit comes at the cost of a reduced system DOF, i.e., $N_a N_P < NM$, which results in decreased performance.

In addition, the non side-looking configuration is very important because the new generation phased-array airborne radars always adopt several antennas mounted in various orientations to scan the moving target in all directions. Similarly, the received data can also be viewed as a series of different components like (11) [2-3]. However, because the Doppler frequency is related both with the look direction and range, the clutter ridge is range-dependent and behaves non-stationary even if the scenario is fully homogenous. Statistical-based methods add preprocessing to improve the stationarity of the training data [10-12]. However, they all need accurate geometry information of the airborne radar, which is difficult to guarantee due to some practical factors such as mechanical vibration and calibration error. Besides, this preprocessing can only compensate the mainlobe response and the sidelobe clutter suppression is limited especially for the short-range case. On the other hand, D3LS is suitable for the non side-looking case by the property of not requiring the training data. However the problem of a reduced system DOF still remains, which will result in performance loss. Focused on the above problems, a new D3 method is proposed in the next section to deal with non-stationary clutter scenario from a different perspective.

## 3. SPACE-TIME SPECTRUM ESTIMATION

Due to the system limitation of the airborne radar, the space-time samples of STAP are insufficient and the slow-moving target is not visible when covered by the surrounding clutter spread. Statistical-based methods utilize the stationary training data to obtain the clutter spectral distribution and design the filter to eliminate the clutter spread in the test cell. However, this stationarity is destroyed in the fast-changing scenario or non side-looking array configuration. However if we can obtain a high-resolution spectrum with only the test cell, both the clutter spread caused by the insufficient samples and the nonstationarity of the training data are effectively eliminated. From this perspective, we have developed a new D3 approach, D3SR, which is based on the technique of sparse representation to obtain a high-resolution spectrum. Unlike current estimators such as Capon and Music [28], D3SR can obtain the high-resolution spectrum with only the test cell. The estimation result includes the possible moving target, the clutter and discrete interference. Similar to that in D3LS, the prior knowledge of SOI is used to extract the clutter and interference distribution from the estimation result so that the CCM and the corresponding adaptive filter are effectively obtained. Next the procedures of D3SR are elaborated.

First discretize the space angle and Doppler frequency axes into $N_s = \rho_s N, N_t = \rho_t M$ grids in the angle-Doppler domain. The parameters $\rho_s, \rho_t$ are the resolution scales along the angle and Doppler axes, respectively. Thus, the received data of the test cell can be written in the matrix form as

$$\mathbf{x} = \sum_{i=1}^{N_s N_t} \alpha_i \cdot \mathbf{\Psi}_i + \mathbf{n} = \mathbf{\Psi}\boldsymbol{\alpha} + \mathbf{n}, \tag{13}$$

where $NM \times N_s N_t$ matrix $\mathbf{\Psi}$ is the overcomplete basis composed of the space-time steering vectors and can be expressed as

$$\mathbf{\Psi} = \left[\boldsymbol{\varphi}(\beta_{s,1}, f_{d,1}), \cdots, \boldsymbol{\varphi}(\beta_{s,N_s}, f_{d,1}), \cdots, \boldsymbol{\varphi}(\beta_{s,1}, f_{d,N_t}), \cdots, \boldsymbol{\varphi}(\beta_{s,N_s}, f_{d,N_t})\right]. \tag{14}$$

The symbols $\beta_{s,i}, 1 \leq i \leq N_s$ and $f_{d,i}, 1 \leq i \leq N_t$ denote the uniformly quantized directions and Doppler frequencies respectively. The vector $\boldsymbol{\alpha}$ stands for the distribution of the test cell in the basis $\mathbf{\Psi}$, (i.e., the space-time spectrum). The basis matrix $\mathbf{\Psi}$ is overcomplete and highly correlated because the resolution scales $\rho_s, \rho_t$ are set greater than one to obtain the high-resolution spectrum. Thus, (13) is ill-posed and has many solutions. However, according to the theory of sparse representation [18-19], the constraint of the sparsity on the actual spectral distribution $\boldsymbol{\alpha}_0$ can help to get rid of this ill-posedness and efficiently solve the equation. The following subsection provides discussion to verify the sparse characteristic of the test cell.

### 3.1 The sparsity of spectral distribution

As stated above, the angle-Doppler domain is discretized into $N_s, N_t$ cells along the angle and Doppler axes, respectively. Each cell in this discretized plane corresponds to a certain space-time steering vector and all of these vectors comprise the overcomplete basis $\mathbf{\Psi}$. Because $\psi = 0$ is a special case, the discussion of sparsity is focused on the general scenario of a non side-looking array. As shown in Fig.3, the clutter ridges of different range cells behave as a set of concentric ellipses. Statistical-based methods using the adjacent range cells cause a widened clutter notch, which suppresses the slow-moving target in the test cell. In the D3 method, the clutter distribution is only one ellipse because no adjacent range cells are needed. Due to the weighting of the radar transmitter, the clutter from the sidelobe is much smaller (below -20dB) than that from the mainlobe. Consequently, the clutter mainly exists in a small area marked with slash cells. Besides the clutter, there are discrete interferers and possible moving target in this plane, which are marked with green triangle and red circle, respectively. Thus, the significant elements only exist in the area of the mainlobe as well as several discrete positions. Compared with the whole discretized plane, the number of cells occupied by these significant elements is quite small. Thus, the received data of the test cell is sparse in the angle-Doppler plane. The only difference between the side-looking and non side-looking cases lies in the shape of the clutter ridge, which is determined by the angle-Doppler dependence. Thus, the significant elements exist in the mainlobe along the straight clutter ridge and in several discrete positions such that the sparsity remains in the side-looking case.

In addition, the illustration of sparsity can also be given in terms of DOF, which is a common metric in the STAP research. As explained in [1-3], the number of the whole cells in the angle-Doppler plane reflects the system DOF. Similarly, the cells occupied by the clutter ridge indicate the clutter rank. Normally, the system gain by the STAP is constituted by two parts. One is the static gain by the radar transmitter via the antenna weighting, which is used to suppress the sidelobe clutter and improve the input SCR. Another is the dynamic gain by the multichannel receiver via the STAP technique. As stated above, because the dynamic gain is obtained mainly in the subspace orthogonal to the clutter, a higher system DOF compared with the fixed clutter rank means a better SCR improvement. Thus the sparsity of the clutter distribution also reflects the SCR improvement provided by the STAP. When a great output SCR is expected, a low-sidelobe antenna and/or sufficient system DOF are necessary so that the area occupied by the clutter ridge is much smaller than the whole discretized angle-Doppler plane. Thus, the sparsity of the STAP received data is universal in the angle-Doppler domain.

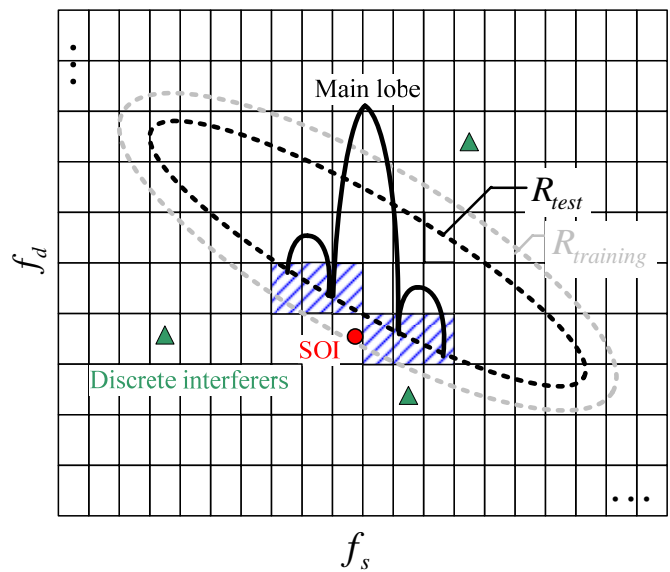

Fig.3 Distribution in the angle-Doppler domain

### 3.2 Spectrum estimation by sparse representation

According to the theory of sparse representation [18-19], when the actual distribution is sparse in a domain, the ill-posed problem in (13) can be efficiently solved using the technique of sparse representation. The basic form of sparse representation is defined as

$$\hat{\boldsymbol{\alpha}} = \arg\min \|\boldsymbol{\alpha}\|_0 \ \ subject \ to \ \ \|\mathbf{x} - \boldsymbol{\Psi}\boldsymbol{\alpha}\|_2 \leq \varepsilon, \tag{15}$$

where $\|\cdot\|_p$ stands for the $L_p$ norm and $\varepsilon$ is the error allowance. However, this problem is a combinatorial problem and NP-hard. To address this difficulty, a number of practical algorithms have been proposed to approximate this sparse solution. One way is to replace the objective function with the $L_1$ norm in (15) [20-21]. It has been proven that this approximation can achieve quite a high performance but also demands a high computational effort especially for large-scale problems.

A series of fast approximation algorithms are also proposed. One effective method, FOCUSS, has been proposed using Lagrange multipliers and can be used to iteratively solve the sparse problem [22-23]. This method uses the weighted $L_2$ norm minimization to make recursive adjustments to the weighting matrix until most elements of the solution are close to zero. The basic form of this recursive method is composed of the following two steps.

$$\mathbf{W}^{(k)} = \text{diag}\left(\boldsymbol{\alpha}^{(k-1)}\right), \tag{16}$$

$$\boldsymbol{\alpha}^{(k)} = \mathbf{W}^{(k)}\left(\boldsymbol{\Psi}\mathbf{W}^{(k)}\right)^{\dagger}\mathbf{x}, \tag{17}$$

where $\text{diag}(\cdot)$ is the operation of diagonalization, $(\mathbf{A})^{\dagger} = \left(\mathbf{A}^H\mathbf{A}\right)^{-1}\mathbf{A}^H$ denotes the pseudoinverse operation of matrix $\mathbf{A}$, $\mathbf{W}^{(k)}$ and $\boldsymbol{\alpha}^{(k)}$ denote the weighting matrix and the corresponding sparse solution at the $kth$ iteration. During the iterations, the FOCUSS method gradually reinforces some of the already prominent entries in $\boldsymbol{\alpha}^{(k)}$ while suppressing the remaining elements until they become close to zero and converge. Prior work by Rao, et al., [23] illustrates that FOCUSS serves as an iterative approximation of the sparse representation using $L_p, 0 \leq p \leq 1$ norm minimization. When the initial value is properly given, FOCUSS can converge to the global optimal value and provide a greater potential of sparse representation and robustness to the noise.

However, the basic FOCUSS has some problems that need improvement. First, there is no inherent mechanism to limit an increase in error from one to the next iteration. For example, if any actual sources are incorrectly eliminated, then it is impossible to retrieve them in the subsequent iteration. Second, the update operation involving the pseudoinverse operation in (17) is carried out in the entire solution space and has a high computation load even though many elements are already close to zero. In the following paragraphs, we address these problems by providing some adjustments to the basic FOCUSS. The corresponding procedures and some practical considerations are given as follows.

a) **Procedures**

1. Obtain the initial value as a low-resolution estimation computed as

$$\alpha_i^{(0)} = \mathbf{s}_i^H \mathbf{x}, 1 \leq i \leq N_s N_t, \tag{18}$$

$$\mathbf{W}^{(0)} = \text{diag}\left(\left[\text{abs}\left(\boldsymbol{\alpha}^{(0)}\right)\right]\right), \tag{19}$$

$$\Gamma = \{i, 1 \leq i \leq N_s N_t\}, \tag{20}$$

where $\mathbf{s}_i$ is the corresponding $ith$ column of basis $\boldsymbol{\Psi}$. The initial vector $\boldsymbol{\alpha}^{(0)}$ is the low-resolution Fourier spectrum of the data. In fact, the initialization does not have to be sparse, otherwise, some potentially elements may be lost.

2. Update the weighting matrix and the estimation result accordingly as

$$\boldsymbol{\alpha}^{(k)}\big|_{\Gamma} = \mathbf{W}^{(k)} \left(\boldsymbol{\Psi}\big|_{\Gamma} \mathbf{W}^{(k)}\right)^{\dagger} \mathbf{x}, \tag{21}$$

where $\boldsymbol{\alpha}^{(k)}\big|_{\Gamma}$ stands for the $\Gamma$ subset of the vector $\boldsymbol{\alpha}^{(k)}$. If current solution $\boldsymbol{\alpha}^{(k)}$ is over-focal, then regenerate a different initial value and carries out a new recursive process 2-4. Else if the solution $\boldsymbol{\alpha}^{(k)}$ is not over-focal, continue to the next step.

3. Update the adaptive subspace using the current solution as

$$\Gamma = \arg\left(\left|\alpha_i^{(k)}\right| \geq Th\right), \ 1 \leq i \leq N_s N_t, \tag{22}$$

where $\alpha_i^{(k)}$ is the $ith$ element of the vector $\boldsymbol{\alpha}^{(k)}$, $Th$ stands for the threshold and the set $\Gamma$ is the adaptive subspace at the $kth$ iteration.

4. Smooth the adaptive subspace set as

$$\mathbf{\Gamma}_{\text{smooth}} = \{\mathbf{\Omega}(i), i \in \mathbf{\Gamma}\}, \tag{23}$$

where

$$\mathbf{\Omega}(i) = \{i, \mathbf{\Lambda}(i)\}, \tag{24}$$

$$\mathbf{\Lambda}(i) = \{u \mid \|r_u - r_i\| \leq d\}, \tag{25}$$

where $r_u, r_i$ stand for the corresponding two-dimensional positions of the $uth, ith$ elements, respectively, $d$ is the distance threshold. Symbol $\mathbf{\Lambda}(i)$ is the two-dimensional neighboring set of the $ith$ position such that the set $\mathbf{\Omega}(i)$ is composed with the $ith$ position and its neighboring set $\mathbf{\Lambda}(i)$. Thus, $\mathbf{\Gamma}_{\text{smooth}}$ is the corresponding smoothed subspace of $\mathbf{\Gamma}$. Next, update the corresponding weighting on the smoothed subspace as

$$\kappa_i = \frac{1}{s_i + 1}\left(\alpha_i^{(k)} + \sum_u \alpha_u^{(k)}\right), i \in \mathbf{\Gamma}_{\text{smooth}}, u \in \mathbf{\Lambda}(i), \tag{26}$$

where $s_i$ stands for the number of neighboring of the $ith$ position, and the weighting matrix can be generated as

$$\mathbf{W}^{(k+1)} = \text{diag}\left(\{\kappa_i, i \in \mathbf{\Gamma}_{\text{smooth}}\}\right), \tag{27}$$

and then the adaptive subspace is updated as

$$\mathbf{\Gamma} = \mathbf{\Gamma}_{\text{smooth}}. \tag{28}$$

5. If the solution is converged as

$$\left|\frac{\boldsymbol{\alpha}^{(k)} - \boldsymbol{\alpha}^{(k-1)}}{\boldsymbol{\alpha}^{(k)}}\right| \leq \varsigma, \tag{29}$$

where $\varsigma$ stands for a small constant, then end the iteration process. Otherwise, repeat the recursive process 2-4.

Unlike the basic implement, adaptive FOCUSS is carried out in a smaller subspace, (i.e., the pseudoinverse operation does not need to incorporate all the columns of the matrix $\mathbf{\Psi}$). Therefore, the adaptive FOCUSS can decrease the computational load while maintaining nearly the same performance as the basic FOCUSS. Additionally, a smoothing operation is needed to keep the recursive process from the error accumulation between iterations and to prevent an over-focal solution. If any actual sources are incorrectly eliminated at the former iteration, it is possible to get them back in the subsequent iterations.

**b) Practical considerations**

As stated above, each iteration requires a pseudoinverse operation. When the matrix $\mathbf{\Psi}|_{\Gamma}\mathbf{W}^{(k)}$ is ill-conditioned, even small noise in the data will result in large change in response. Here we use a common regularization technique called truncated singular value decomposition (TSVD) to improve the numerical stability, which is detailed in [22]. In addition, during the iteration, we need to determine whether current sparse solution is over-focal. This judgment is difficult because the actual sparse solution is unknown in advance. Here we simply compare the current sparse solution with the Fourier spectrum, which is low-resolution but unbiased. If the envelope of the estimated solution is quite different from the Fourier spectrum, this solution may be over-focal. Because FOCUSS is essentially an iterative algorithm to obtain the sparse solution, it may be trapped in some local minima. Thus, a smoothing operation is necessary to prevent the estimation error accumulation between iterations. The smoothing is only confined to a neighboring region around each significant cell in the angle-Doppler plane. The smoothing region should not be too large, otherwise, it may cause a over-smoothed solution, which does not become sparse between iterations. In our studies, the neighboring region is defined as the nearest eight cells around the center in the discretized angle-Doppler plane.

### 3.3 Target detection

After obtaining the spectral distribution of the test cell with high accuracy, it is possible to make a direct amplitude detection in the actual SOI channel. However, because the columns in the overcomplete dictionary are highly correlated, the estimation using sparse representation only guarantees that the significant elements are accurately recovered [24-25]. Consequently, when we aim to locate a moving target from the surrounding strong clutter in the airborne radar scenario, the estimation of the target amplitude may be not reliable. To avoid this, we only extract the significant clutter and interference components from the estimation result and design the adaptive filter to suppress them. Here, D3SR follows the similar concept of D3LS to extract the clutter distribution with the assumed SOI [15-16]. However, D3SR extracts the clutter from the estimated result in the angle-Doppler domain, as opposed to D3LS, which extracts the clutter in the direct data domain. After obtaining the spectral distribution of the clutter and interference, the adaptive filter can be designed to effectively suppress both of these elements. The details of the D3SR method are elaborated as follows.

1. Determine the SOI area using the priori knowledge as

$$\Gamma(\theta_s, f_d) = \left[ p_{1,1}, \cdots p_{1,M_{SOI}}, \cdots, p_{N_{SOI},M_{SOI}} \right] \tag{30}$$

where $p_{i,j}, 1 \leq i \leq N_{SOI}, 1 \leq j \leq M_{SOI}$ stand for the possible indexes of the SOI area in the discretized angle-Doppler plane. The extent of the SOI area, (i.e., $N_{SOI}, M_{SOI}$) reflects the uncertainty along angle and Doppler axes.

2. Once the SOI area is determined, the clutter distribution can be extracted from the spectrum estimation, and the corresponding CCM estimation can be given as

$$\hat{\mathbf{R}}_{SR} = \sum_i |\hat{\alpha}_i|^2 \boldsymbol{\varphi}(\theta_{s,i}, f_{d,i}) \boldsymbol{\varphi}(\theta_{s,i}, f_{d,i})^H + \beta_L \mathbf{I}, \quad i \notin \Gamma(\theta_s, f_d), \tag{31}$$

where $|\hat{\alpha}_i|^2$ is the space-time spectrum for the $ith$ clutter scatter using the sparse representation, $\beta_L$ is a small loading factor. Thus, $\hat{\mathbf{R}}_{SR}$ corresponds to the distribution characteristics of clutter and interference in the test cell.

3. The space-time adaptive filter can then be given as

$$\mathbf{w}_{SR} = \mu \hat{\mathbf{R}}_{SR}^{-1} \mathbf{s}, \tag{32}$$

where $\mathbf{s}$ stands for the $NM \times 1$ space-time steering vector of the moving target.

After the above procedures, the adaptive filter can be effectively built using the spectrum estimation of sparse representation and the assumed SOI. D3SR obtains an accurate CCM estimation only with the test cell such that the problem of lacking stationary training data is well solved. However, it is still necessary to calculate the CCM inversion in (32), which has a high computation load. When the clutter rank is low, there are several iterative approaches to approximate the CCM inversion [29]. However, these fast inversion methods are not covered in this paper and our attention is mainly focused on an accurate CCM estimation with only the test cell.

## 4. EXPERIMENTAL RESULTS

In the simulation section, the scenarios of both side-looking and non side-looking airborne linear arrays are considered. The basic parameters are given in Table I.

Table I System parameters

| Parameter | Symbol | Value |
|---|---|---|
| Number of sensors | $N$ | 12 |
| Number of pulses | $M$ | 12 |
| Platform velocity | $v$ | 300m/s |
| Pulse repetition interval | $PRI$ | 0.25ms |
| Sample rate | $f_s$ | 5Mhz |
| Radar wavelength | $\lambda$ | 0.3m |
| Inter-sensor spacing | $d$ | 0.15m |
| Platform height | $H$ | 3000m |
| Signal-to-clutter ratio | $SCR$ | -30dB |

In the non side-looking case, the clutter is range-dependent and behaves non-stationary by the intrinsic array configuration. Therefore no interference but the range-dependent clutter is considered in this case. The clutter is uniformly distributed between the directions $90° \sim 160°$ and the crab angle $\psi$ is set as $45°$. Here only the front-lobe clutter responses are considered. The range samples begin from the vertical height $H$ with 100 range cells, which is a typical short-range case.

In the side-looking case, the clutter ridge is range-independent. Therefore both the discrete interference and clutter is considered to generate the nonstationarity. The clutter is uniformly distributed between the directions $20° \sim 60°$ and is contained in both the training data and test cell. The discrete interferers only appear in the test cell and the parameters are given in Table II.

Table 2 Parameters of discrete interference

| Angle-of-arrival | $-60°$ | $-40°$ | $-20°$ | $40°$ | $60°$ |
|---|---|---|---|---|---|
| Normalized Doppler frequency | 0 | 0.1 | 0.2 | 0.1 | -0.4 |

The resolution scales in both two cases are set as $\rho_s = 6, \rho_t = 6$ to obtain the high-resolution spectrum. As with the sparse representation, the convex optimization package cvx [30] is employed as the $L_1$ norm minimization tool. The adaptive FOCUSS takes the TSVD regulation in accordance with the noise level $\delta^2$, and the neighboring threshold $d$ is set as $\sqrt{2}$. The SOI is marked as a circle while the discrete interferers and clutter are marked as triangle and ellipse symbols, respectively, in the following figures. The spectrum estimation using both $L_1$ norm minimization and adaptive FOCUSS is first given to verify the advantages of sparse representation. Then the output spectrums using different STAP methods are compared. Finally, the performance of the range output and MDV is also tested.

**4.1 Spectrum estimation using sparse representation**

In the side-looking case, the moving target is located at the $14th$ range cell, coming from the direction $15°$ with a normalized Doppler frequency 0.3. Fig.5 (a) shows the input spectrum using the Fourier transform. Figs.5 (b)-(c) show the spectrum estimation using $L_1$ norm minimization and adaptive FOCUSS, respectively. The moving target is submerged in the surrounding clutter and interference environment and not obviously seen in the input spectrum. However, both the $L_1$ norm minimization and the adaptive FOCUSS can obtain the high-resolution spectrum estimation. Thus, the sparse representation using the both two implements can decrease the clutter and interference spread so that the moving target is visible in the estimated spectrums. However, because the $L_1$ norm minimization tends to express the data with fewer space-time steering vectors, the continuous clutter ridge converges into several discrete scatters and may lose some actual scatters. In contrast to $L_1$ norm minimization, the adaptive FOCUSS is essentially an iterative approximation of $L_p, 0 \leq p \leq 1$ norm minimization. Thus, as stated above, it often has a greater potential of sparse representation and the spectral estimation matches better with the actual clutter scenario.

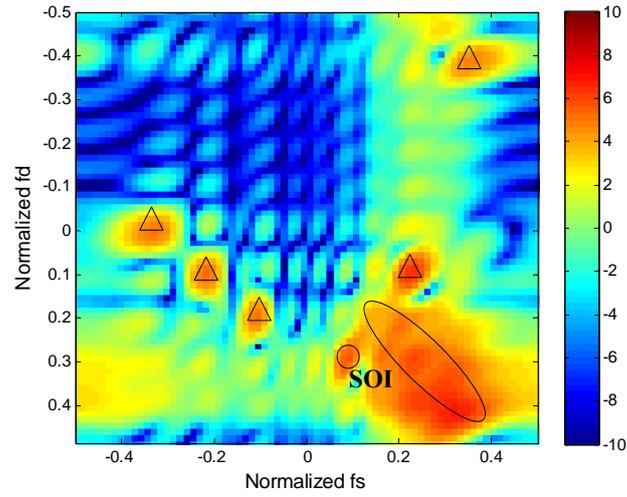

Fig.5 (a) Input spectrum (dB) in side-looking case

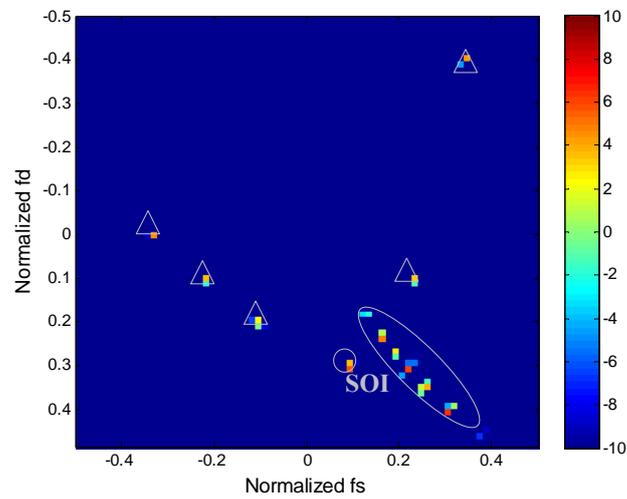

Fig.5 (b) Spectrum estimation (dB) using $L_1$ norm in side-looking case

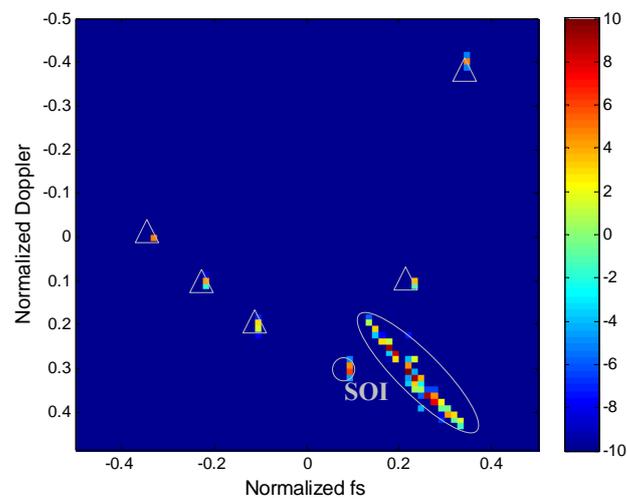

Fig.5 (c) Spectrum estimation (dB) using adaptive FOCUSS in side-looking case

Figs.6 (a)-(c) gives the parallel results in the non side-looking case. The moving target is located at the $14th$ range cell, coming from the direction $135°$ with a normalized Doppler frequency 0.25. Similar to the above simulation, the spectrum estimation using both $L_1$ norm minimization and adaptive FOCUSS obtain the high-resolution spectrum and can decrease the clutter spread. Unlike the non side-looking case, here, the clutter ridge behaves part of an ellipse instead of a straight line. However, it is notable that the target in the estimation result is detectable but its amplitude is not accurate in both side-looking and non side-looking cases. Thus the moving target may not be directly detected in the estimation spectrums, which also illustrates the necessity for the adaptive filter in the subsection 3.3.

In addition, the dimension of the subspace in adaptive FOCUSS keeps decreasing during the iterations, while the $L_1$ norm minimization always carries out the optimization in the whole space. Thus, the adaptive FOCUSS can obtain better spectrum estimation and has a smaller computation load than $L_1$ norm minimization. In the following parts, the D3SR filter adopts the estimation result of the adaptive FOCUSS.

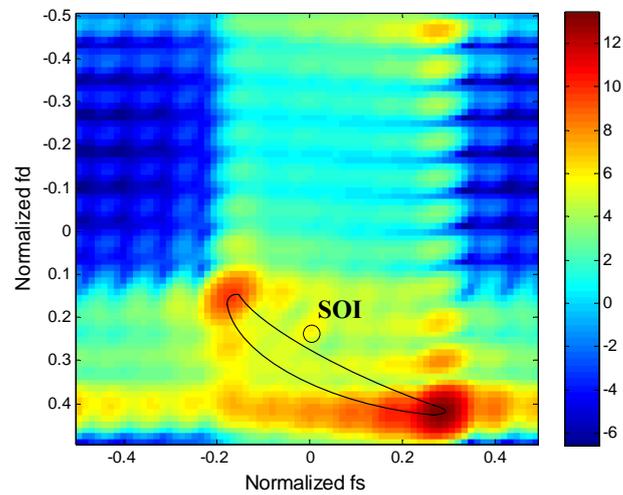

Fig.6 (a) Input spectrum (dB) in non side-looking case

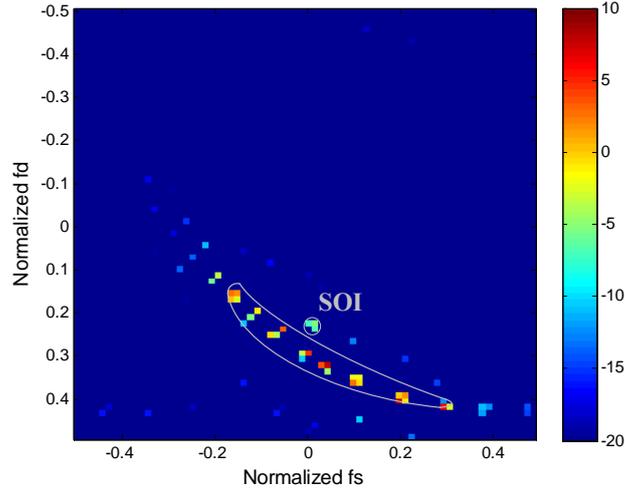

Fig.6 (b) Spectrum estimation (dB) using $L_1$ norm in non side-looking case

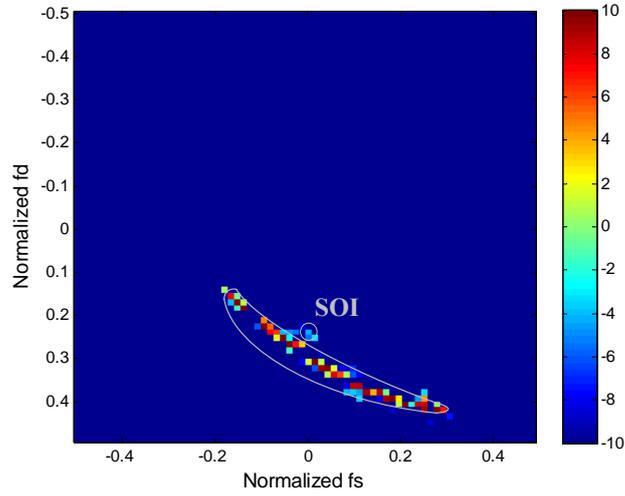

Fig.6 (c) Spectrum estimation (dB) using adaptive FOCUSS in non side-looking case

**4.2 Output performance**

Once the spectral distribution of the test cell is obtained, the clutter and interference spectral characteristics can be extracted using the assumed SOI and the adaptive filter can be effectively built. Figs.7 (a)-(c) give the output spectrums using different STAP methods in the side-looking case. Here LSMI is also adopted as a reference to illustrate the problem of statistical-based methods. The number of training data is $L = 2NM$ for LSMI, while no training data but only the test cell is for both the D3LS and D3SR methods. In the D3LS method, we use $N_a = 8, N_p = 8$ with the forward type [16]. The results show that the interference still exists after the LSMI processing because it is not contained in the training data. However, both the D3LS and D3SR methods can effectively suppress the clutter and interference. The difference between the D3LS and D3SR methods is that D3SR provides an effective filter without a reduced system DOF so that it can achieve better output SCR performance.

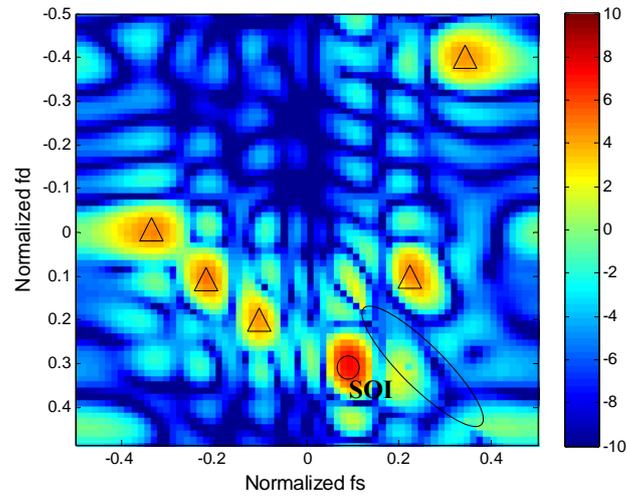

Fig.7 (a) LSMI output spectrum (dB) in side-looking case

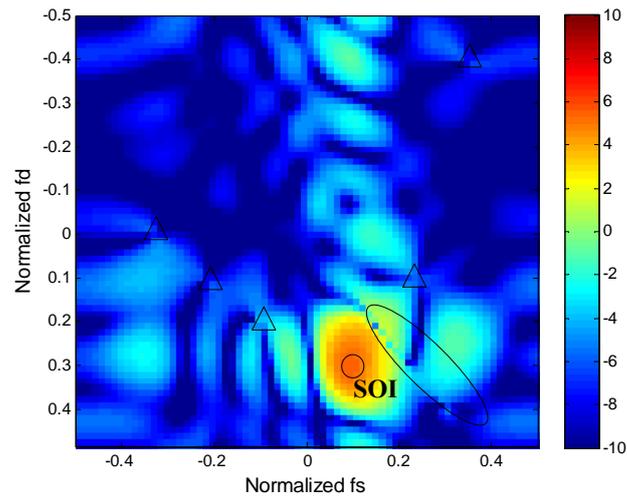

Fig.7 (b) D3LS output spectrum (dB) in side-looking case

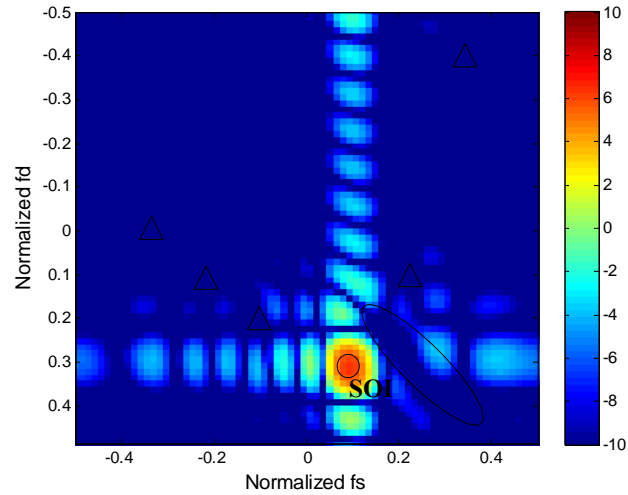

Fig.7 (c) D3SR output spectrum (dB) in side-looking case

Figs.8 (a)-(c) give the parallel results in the non side-looking case, where only range-dependent clutter is considered to generate the non-stationary scenario. Because LSMI needs adjacent range cells to estimate the CCM, the range-dependence of the clutter ridge will cause an inappropriate filter and suppress the slow-moving target. On the other hand, because no training data but only the test cell is acquired in both the D3LS and D3SR methods, they can avoid the nonstationarity of the training data and effectively suppress the clutter to make the target visible. Similar to the side-looking case, the output spectrum of D3SR is better than that of D3LS because there is no loss in the system DOF when using the D3SR method. Further explanation will be provided in subsection 4.4.

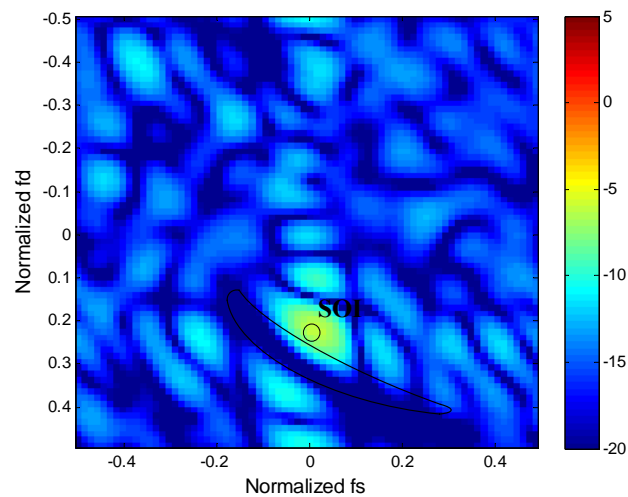

Fig.8 (a) LSMI output spectrum (dB) in non side-looking case

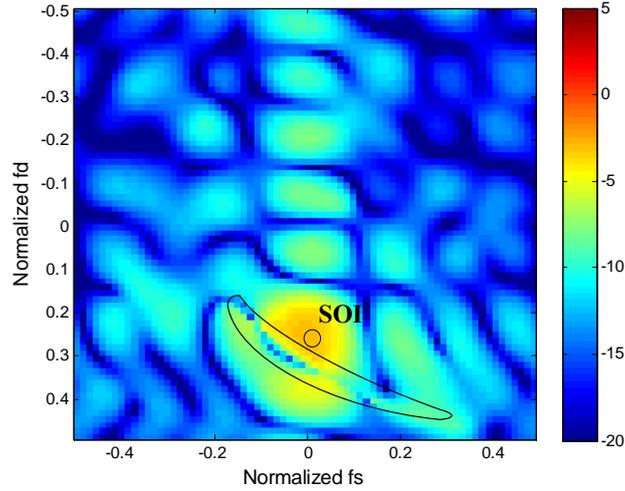

Fig.8 (b) D3LS output spectrum (dB) in non side-looking case

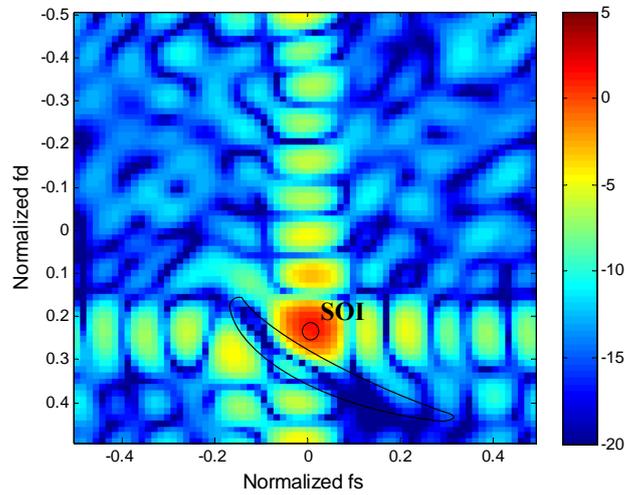

Fig.8 (b) D3SR output spectrum (dB) in non side-looking case

**4.3 Target detection along range cell**

In this subsection, the target detection along range cells in both side-looking and non side-looking cases is tested. In the side-looking case, the parameters of the moving target and clutter distribution keep the same as the above simulation. The discrete interferer located at the $30, 60th$ range cells comes from the direction $-20°$, with a normalized Doppler frequency 0.3. Because the discrete interferer and the moving target have the same Doppler frequency, the interference sidelobe will impact the target detection. Fig.9 (a) shows the corresponding range output in the SOI channel with different STAP algorithms. As shown in the "Non" curve, the actual target is completely obscured by the clutter and interference sidelobe so that it can hardly be detected without the adaptive processing. The LSMI method uses the adjacent range cells to estimate the CCM and construct the adaptive filter. Because the training data does not contain any information about the discrete interferer, the interferer still remains in the $30, 60th$ range cells even after the LSMI filter. Consequently, the interference residual impacts the target detection and may cause false alarms along the range cells. Conversely, the D3 methods including both D3LS and D3SR can effectively suppress the clutter and the interference. Thus, the output SCR has been greatly improved so that the moving target can be obviously seen. Because D3SR has a higher system DOF to design the adaptive filter, it has less clutter residual along the range cells and owns a better target detection performance.

Fig.9 (b) shows the parallel results in the non side-looking case. The simulation scenario keeps the same as before, where no interference but only the range-dependent clutter is considered. Because LSMI utilizes the adjacent range cells, it forms a widened clutter notch and suppresses the slow-moving target. Thus, the target is not visible in the range output after the LSMI filter. Alternatively, both D3LS and D3SR avoids the requirement of the training data, thus they can solve the problem of nonstationarity and effectively suppress the range-dependent clutter. Similarly, because there is no loss of system DOF, there is less clutter residual along range cells after the D3SR filter.

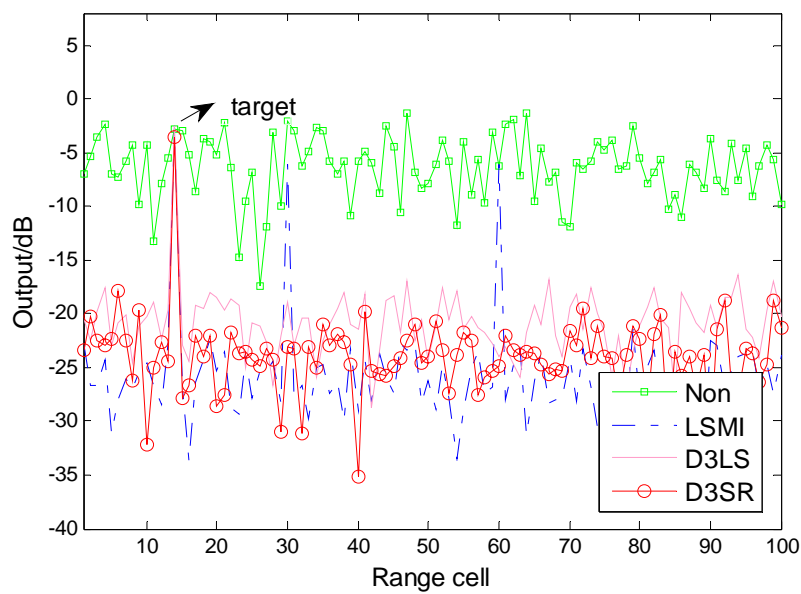

Fig.9 (a) Range output (dB) in side-looking case

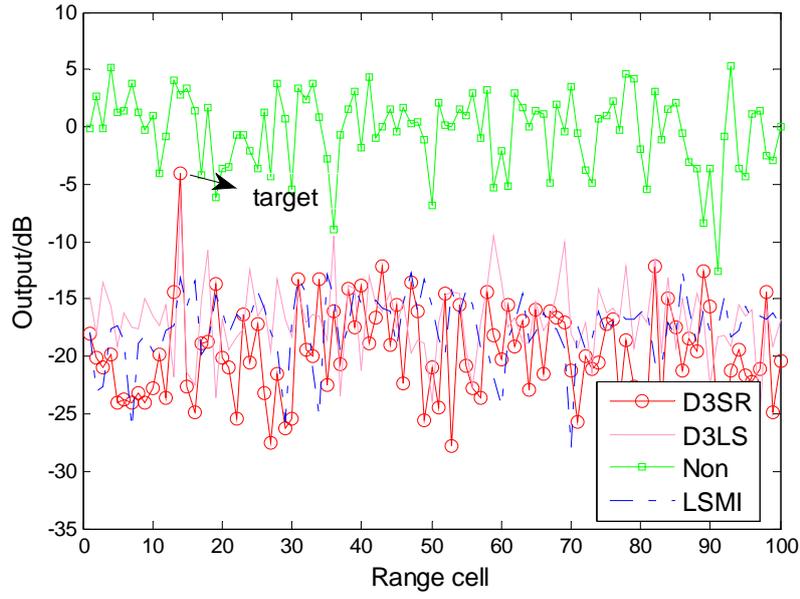

Fig.9 (b) Range output (dB) in non side-looking case

**4.3 MDV performance**

In the airborne radar system, it is quite important to detect and locate a slow-moving target in the presence of the surrounding clutter [1-2]. A common metric of this performance is the MDV for an acceptable output SCR. Thus the output SCR with a different target velocity is considered in this subsection. The simulation parameters are kept the same in Table I, while the moving target is coming from the direction $30°$ and $135°$ in side-looking and non side-looking cases, respectively. Traditional measurement needs statistical information to evaluate this metric, which is unknown in the D3 case. Thus, we simply calculate the output SCR as

$$SCR_{out} = \frac{\left|\mathbf{w}^H \mathbf{x}_t\right|^2}{\left|\mathbf{w}^H \left(\mathbf{x}_c + \mathbf{x}_d + \mathbf{n}\right)\right|^2}, \qquad (33)$$

where $\mathbf{w}$ is the adaptive filter given by D3LS or D3SR, $\mathbf{s}$ stands for the space-time steering vector of the moving target. Next, 100 Monte Carlo simulations are carried out to obtain average performance. Figs.10 (a) and (b) give the output SCR along Doppler axis in both side-looking and non side-looking cases. Because the SCR improvement is mostly achieved in the subspace orthogonal to the clutter, both D3LS and D3SR suffer considerable degradation near the clutter notch, no matter what size the total space (i.e., system DOF) is. When the target is away from the clutter notch, it is possible to distinguish the clutter and the slow-moving target. In this case,

D3SR can provide a better output SCR than D3LS, which is owing to the characteristics of full system DOF. Thus it can provide a narrower clutter notch and owns better SCR improvement in the pass-band area than D3LS. Thus D3SR is more effective against both the range-dependent clutter and discrete interference, and has a great potential in the non-stationary scenario.

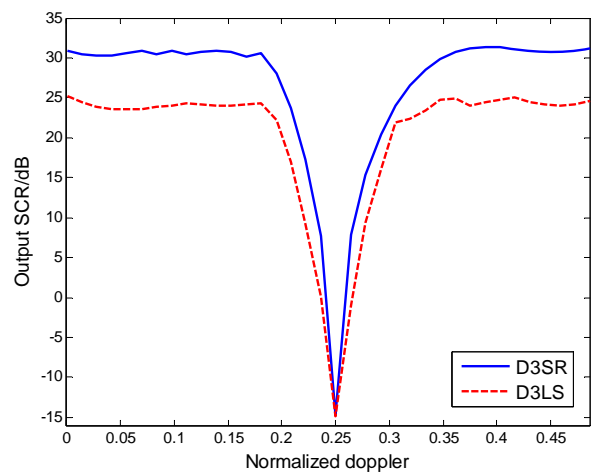

Fig.10 (a) Output SCR along Doppler axis in side-looking case

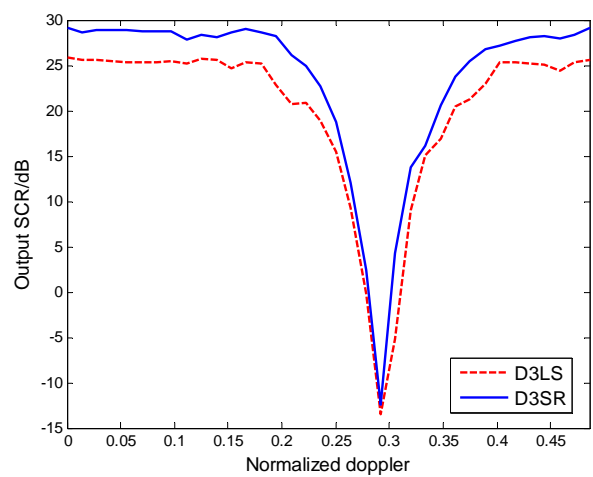

Fig.10 (b) Output SCR along Doppler axis in non side-looking case

## 5. CONCLUSION

In this paper, we have analyzed the sparsity of the spectral distribution in the angle-Doppler domain and proposed a new D3 method to deal with the non-stationary clutter scenario in both the side-looking and non side-looking cases. Our proposed D3SR method can obtain high-resolution spectrum using sparse representation such as the $L_1$ norm minimization or the adaptive FOCUSS. Based on this, D3SR can obtain an accurate CCM estimation of the test cell and provides an effective adaptive filter with full of system DOF. Thus, it achieves better performance of the output SCR and MDV.

The following are some considerations for further research. First, the current overcomplete dictionary $\Psi$ is fixed in sparse representation. However, due to the practical nonideal factors such as clutter internal motion and/or channel mismatch, this predefined overcomplete dictionary does not match with the actual data and the corresponding sparsity may decrease. Therefore, solving the sparse representation problem where both the overcomplete dictionary and actual sources are unknown seem to be quite important. Second, the proposed D3SR method can be conveniently extended to a more general configuration that is deployed with the bistatic radar and/or conformal array where the clutter ridge also behaves range-dependent. However, in these circumstances, the dictionary and the corresponding sparsity need to be reconsidered. Adaptive mechanisms are also necessary in the sparse representation to guarantee a good estimation result.